\documentclass[prl,letterpaper,twocolumn,preprintnumbers,nofootinbib]{revtex4}
\usepackage[letterpaper, hdivide={1.91cm,,1.165cm}, vdivide={1.83cm,,3.1cm}]{geometry}

\usepackage{amsmath,amssymb}
\usepackage{graphicx}
\usepackage{color}
\usepackage{hyperref}
\usepackage{microtype} 

\def \vec#1{{\boldsymbol{#1}}}

\begin{document}

\title{Revisiting the connection of baryon number, lepton number, and operator dimension}

\author{Julian Heeck}
\email[Email: ]{heeck@virginia.edu}
\thanks{ORCID: \href{https://orcid.org/0000-0003-2653-5962}{0000-0003-2653-5962}.}
\affiliation{Department of Physics, University of Virginia,
Charlottesville, Virginia 22904-4714, USA}

\author{Diana Sokhashvili}
\email[Email: ]{bkq5jz@virginia.edu}
\thanks{ORCID: \href{https://orcid.org/0009-0002-0428-6264}{0009-0002-0428-6264}.}
\affiliation{Department of Physics, University of Virginia,
Charlottesville, Virginia 22904-4714, USA}

\hypersetup{
pdftitle={Revisiting the connection of baryon number, lepton number, and operator dimension},   
pdfauthor={Julian Heeck, Diana Sokhashvili}
}

\begin{abstract}
    The effects of heavy new particles beyond the Standard Model can be conveniently captured through higher-dimensional effective operators. As noted long ago by Weinberg, the amount of baryon and lepton number an operator can carry is intricately connected to its mass dimension. We derive an improved inequality for this connection and compare it to explicit operator constructions up to mass dimension 25. For the effective field theory of Standard Model plus right-handed neutrinos, our relationship is even an equality up to high mass dimension.
\end{abstract}

\maketitle

The Standard Model (SM) has proven remarkably successful in describing particle interactions up to high energies. The non-observation of new particles at the Large Hadron Collider so far indicates that any particles beyond the SM must be very heavy, at least if they carry charges under the SM gauge group, $SU(3)_C\times SU(2)_L\times U(1)_Y$.  This in turn allows us to integrate-out such heavy states and capture their effects in an effective field theory that extends the SM Lagrangian by higher-dimensional operators. This Standard Model Effective Field Theory (SMEFT) conveniently parametrizes any and all effects of heavy new particles, organized essentially by the number of new-particle propagators in underlying Feynman diagrams.

Of particular interest are SMEFT operators that break the global SM symmetries, baryon and lepton numbers, as they typically lead to novel and background-free signatures, allowing us to probe those kinds of operators very precisely~\cite{Heeck:2019kgr,Dev:2022jbf,FileviezPerez:2022ypk,Broussard:2025opd}. As shown by Weinberg~\cite{Weinberg:1979sa,Weinberg:1980bf} and others~\cite{Weldon:1980gi,Rao:1983sd,Degrande:2012wf, deGouvea:2014lva, Henning:2015alf,Heeck:2019kgr}, the number of baryons, $\Delta B$, and leptons, $\Delta L$, that an effective operator carries varies distinctly with the operator's mass dimension~$d$. For example, $|\Delta L|=2$, $\Delta B=0$ operators appear at any odd mass dimension $d \geq 5$, while $\Delta B=\Delta L\neq 0$ operators arise at even mass dimension $d\geq 6$.
A careful study of the connection between $\Delta B$, $\Delta L$, and operator dimension $d$ was performed by Kobach~\cite{Kobach:2016ami}, which in particular revealed the following inequality for the minimal mass dimension $d_\text{min}$ of an operator with $(\Delta B,\Delta L)$:
\begin{align}
    d_\text{min} \geq \frac{9}{2}|\Delta B|+\frac{3}{2} |\Delta L| \,.
    \label{eq:Kobach}
\end{align}
This inequality is easy to understand: baryon number is only carried by quarks, whose $SU(3)_C$ color charge forces them to come in multiples of triplets $qqq$, each of which carrying $\Delta B=1$ and $d=9/2$; lepton number is only carried by, well, leptons, which contribute an extra $d=3/2$ for every $\Delta L$, leading to Eq.~\eqref{eq:Kobach}. The equality can only be realized in operators without derivatives or bosonic fields, as they would increase the mass dimension without affecting $\Delta B$ or $\Delta L$. Angular momentum further requires $|\Delta B| + |\Delta L|$ to be an even integer, see Ref.~\cite{Kobach:2016ami} for further relationships along these lines.

\begin{table}[tb]
\centering
{\renewcommand{\arraystretch}{1.2}
\begin{tabular}{c c c c c} 
 \hline
 field & chirality & copies & $SU(3)_C\times SU(2)_L\times U(1)_Y$   & $F$\\ [0.5ex] 
 \hline\hline
 $Q$ & left & 3 & $\left(\vec{3},\vec{2},\tfrac16\right)$ & $+$\\
 $u$ & right & 3 & $\left(\vec{3},\vec{1},\tfrac23\right)$& $+$\\
 $d$ & right & 3 & $\left(\vec{3},\vec{1},-\tfrac13\right)$ & $+$\\
 $L$ & left & 3 & $\left(\vec{1},\vec{2},-\tfrac12\right)$ & $+$\\
 $\ell$ & right & 3 & $\left(\vec{1},\vec{1},-1\right)$ & $+$\\
 $\nu$ & right & 3 & $\left(\vec{1},\vec{1},0\right)$ & $+$\\
 $H$ & scalar & 1 & $\left(\vec{1},\vec{2},\tfrac12\right)$ & $-$\\
 \hline
 
  $Q^C$ & right & 3 & $\left(\bar{\vec{3}},\vec{2},-\tfrac16\right)$ & $-$\\
 $u^C$ & left & 3 & $\left(\bar{\vec{3}},\vec{1},-\tfrac23\right)$& $-$\\
 $d^C$ & left & 3 & $\left(\bar{\vec{3}},\vec{1},\tfrac13\right)$ & $-$\\
 $L^C$ & right & 3 & $\left(\vec{1},\vec{2},\tfrac12\right)$ & $-$\\
 $\ell^C$ & left & 3 & $\left(\vec{1},\vec{1},1\right)$ & $-$\\
 $\nu^C$ & left & 3 & $\left(\vec{1},\vec{1},0\right)$ & $-$\\
 $H^C$ & scalar & 1 & $\left(\vec{1},\vec{2},-\tfrac12\right)$ & $-$\\
 \hline
 $D_\mu$ & vector & / & / & $-$\\
 \hline
\end{tabular}
}
\caption{SM fields (plus right-handed neutrinos $\nu$) and quantum numbers; hypercharge $Y$ is related to electric charge $Q$ via $Q = Y + T_3$, with diagonal $SU(2)_L$ generator $T_3$. The last column lists Weinberg's $F$ parity~\cite{Weinberg:1980bf}.}
\label{tab:fields}
\end{table}

Improving Eq.~\eqref{eq:Kobach} requires the imposition of gauge and Lorentz invariance. Kobach has shown that $U(1)_Y$ invariance is a sufficient condition for gauge invariance~\cite{Kobach:2016ami}, albeit difficult to implement in Eq.~\eqref{eq:Kobach}.
Here, we note that a weaker condition has been identified by Weinberg in the form of $F$ parity~\cite{Weinberg:1980bf}, which is multiplicatively conserved due to Lorentz invariance and $SU(2)_L$ and has a remarkably simple form: SM fermions are \textit{even} under $F$ parity, anti-fermions \textit{odd}, and bosons and derivatives are \textit{odd}, too, see Tab.~\ref{tab:fields}.

Let us focus on operators with positive odd $\Delta B$ for now, which then has even $F$ parity but requires an odd number of leptonic fields to form a Lorentz singlet. An odd number of \textit{leptons}, i.e.~positive odd $\Delta L$, gives an even-$F$ operator that is hence allowed, while an odd number of \textit{anti-leptons}, i.e.~negative odd $\Delta L$, would lead to an odd-$F$ operator, forcing us to introduce an odd number of additional Higgs doublets or derivatives to conserve $F$ parity, increasing the operator's mass dimension by at least one. More generally, any operator with negative odd $\Delta B \cdot \Delta L$ is odd under $F$ parity and requires an additional Higgs or derivative. Using the \textit{indicator function} $\mathbf{1}_A(x)$, $F$ parity conservation then modifies Kobach's inequality compactly to
\begin{align}
    d_\text{min} \geq \frac{9}{2}|\Delta B|+\frac{3}{2}|\Delta L|+ \mathbf{1}_{1-2\mathbb{N}}(\Delta B\Delta L) \,,
    \label{eq:better_inequality}
\end{align}
i.e.~adding $+1$ if $\Delta B \Delta L$ is negative and odd. This improved inequality still does not fully incorporate gauge invariance of the underlying operator, but captures part of those constraints.
Together with the rule that $\Delta B + \Delta L $ has to be even, this gives a simple estimate for the lowest operator dimension with a given $(\Delta B,\Delta L)$.
Additional operators with the same $(\Delta B,\Delta L)$ can arise at $d_\text{min}+2$, $d_\text{min}+4$,  etc, simply because one can always multiply the operator by the singlet $|H|^2$ or insert two derivatives $D_\mu D^\mu$. $F$~parity \textit{forbids} the same $(\Delta B,\Delta L)$ to arise at $d_\text{min}+$odd number~\cite{Kobach:2016ami}.

It is worth noting that the same inequality~\eqref{eq:better_inequality} holds in the $\nu$SMEFT, i.e.~the SMEFT extended by three copies of SM-singlet right-handed neutrinos $\nu$, see Tab.~\ref{tab:fields}, a popular generalization both due to the increased structural symmetry of the model and the popular role right-handed neutrinos play to provide neutrino masses. As long as the $\nu$ carry the same lepton number and $F$ parity as the other leptons, Eq.~\eqref{eq:better_inequality} remains valid.

\begin{figure*}[!phtb]
\centering
\includegraphics[width=0.94\textwidth]{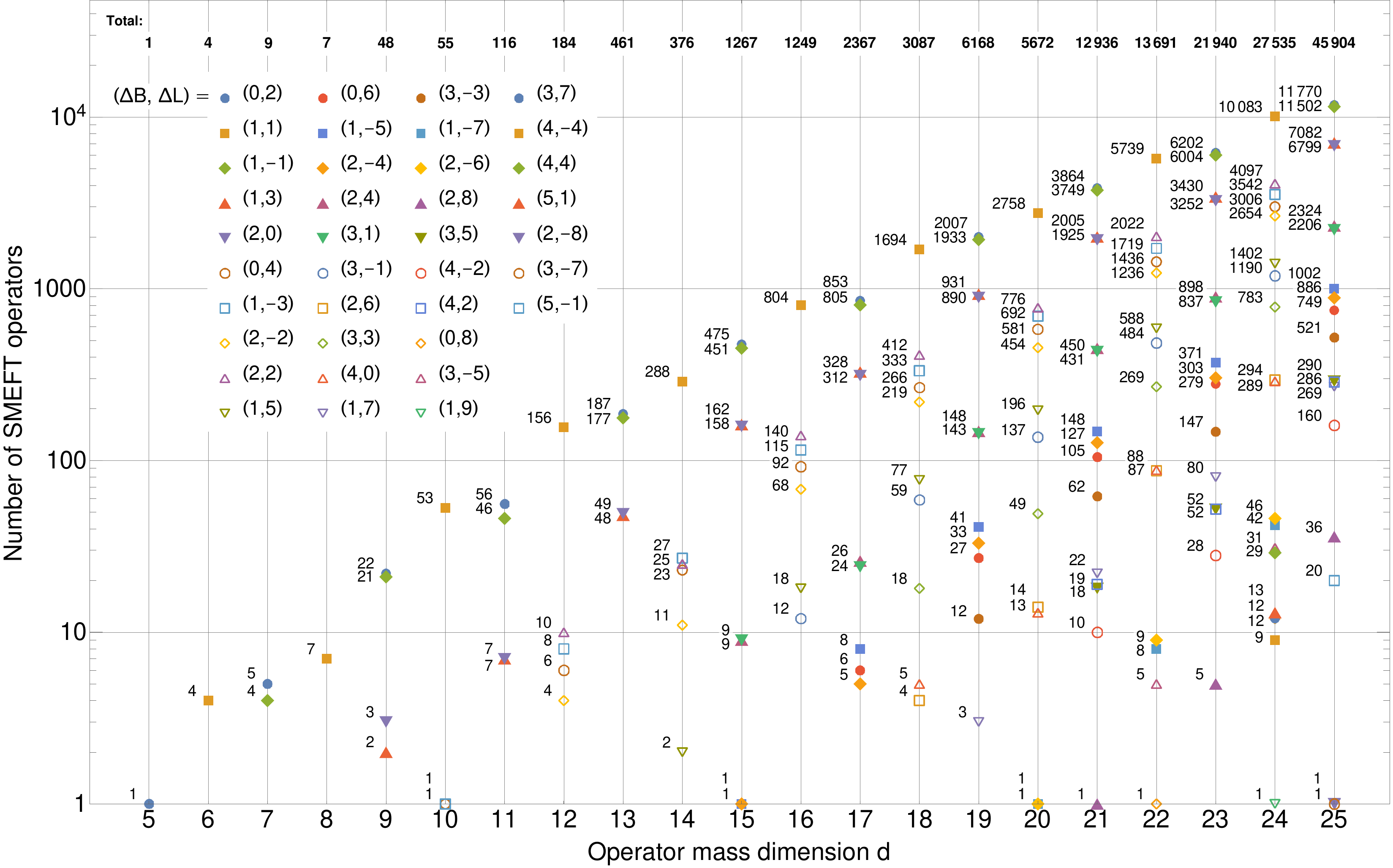}
\caption{Number of non-derivative SMEFT operators in each mass dimension  $d$ grouped by their $(\Delta B,\Delta L)$.
\label{fig:number_of_operators}}

\bigskip

\includegraphics[width=0.94\textwidth]{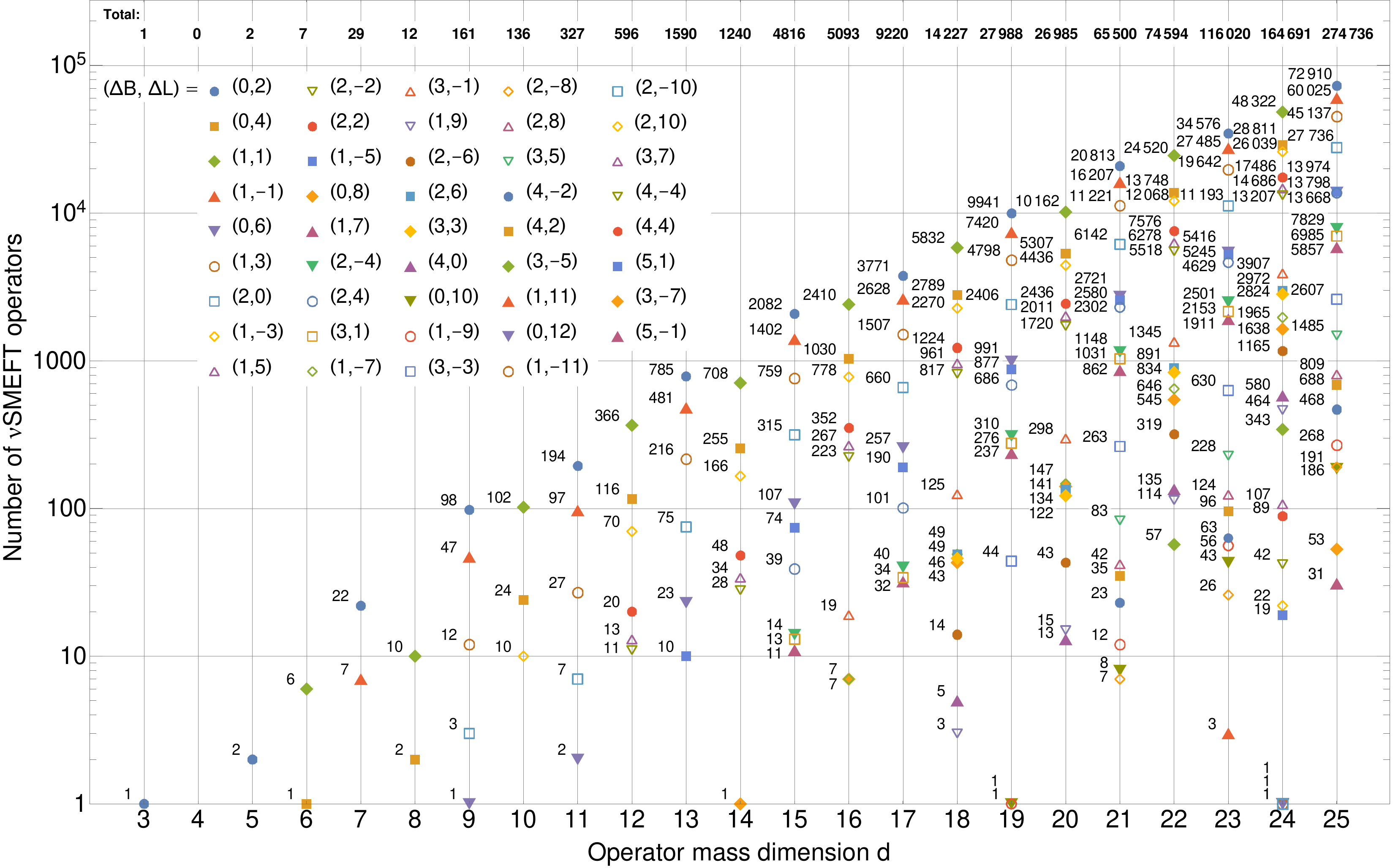}
\caption{Same as Fig.~\ref{fig:number_of_operators} but for $\nu$SMEFT instead of SMEFT.
\label{fig:number_of_nuSMEFT_operators}}
\end{figure*}
 
\begin{figure*}[!ht]
\centering
\includegraphics[width=0.98\textwidth]{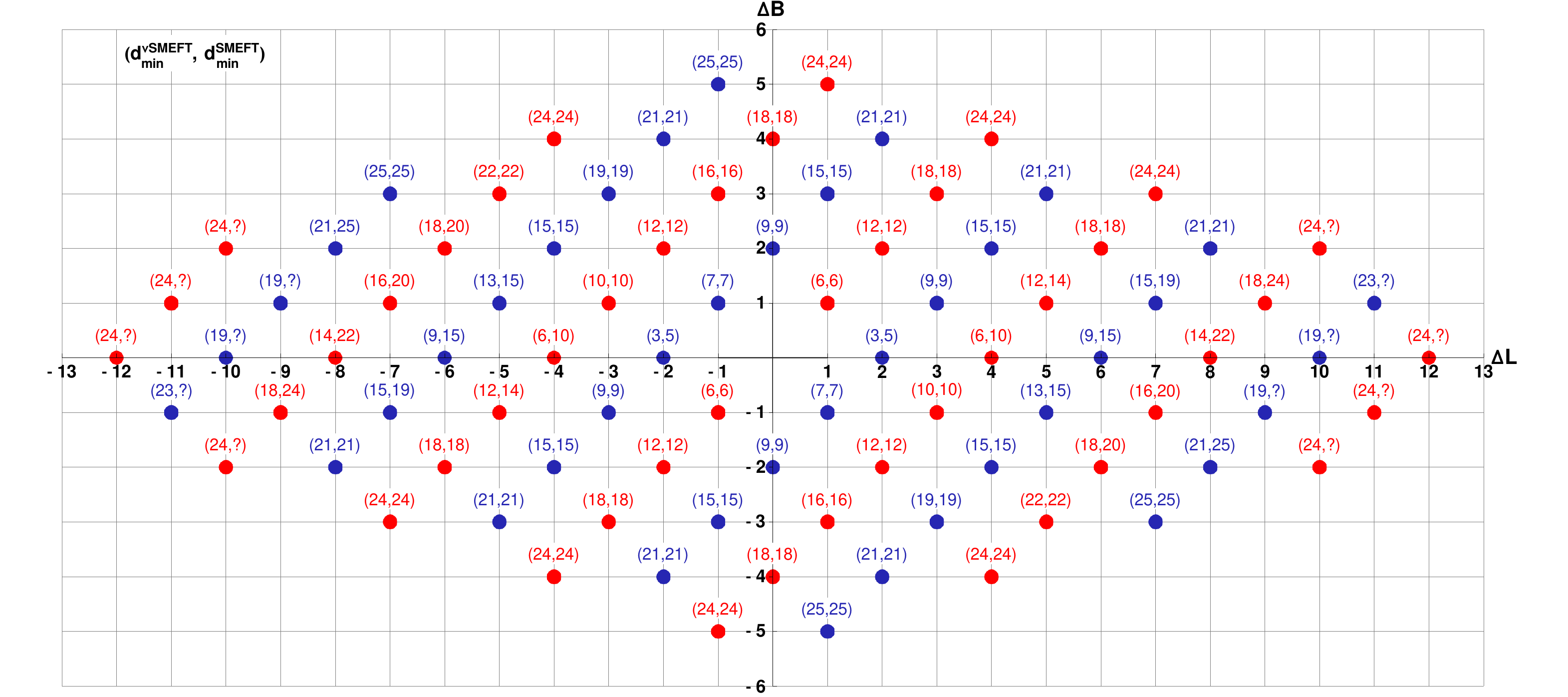}
\caption{($\nu$)SMEFT operators up to mass dimension $d=25$ grouped by their baryon
and lepton number. The tuples $(d_\text{min}^{\nu\text{SMEFT}},d_\text{min}^{\text{SMEFT}})$ above each allowed point indicate the  smallest mass dimension at which the corresponding $\Delta B$ and $\Delta L$ arise. Other operators with the same $\Delta B$ \& $\Delta L$ simply occur at $d_{\rm{min}}+$even number.
For example, $\Delta B = - \Delta L = 1$ operators exist at $d = 7,9,11,13,15,17,\ldots = d_{\rm{min}} + 2\mathbb{N}_0$.
A question mark indicates that $d_\text{min}^{\text{SMEFT}} > 25$, i.e.~above our \texttt{Sym2Int} cutoff.
\label{fig:dBdL}}
\end{figure*}

To test Eq.~\eqref{eq:better_inequality}, we used \texttt{Sym2Int}~\cite{Fonseca:2017lem,Fonseca:2019yya} to explicitly construct SMEFT and $\nu$SMEFT operators up to mass dimension $d=25$.
Reaching this high mass dimension required several days of computing on a high-memory cluster and restriction to non-derivative operators. We also checked derivative operators up to $d=17$ but found no qualitative difference, as expected since derivatives act similarly to $H$ in this context. 
To get a sense of the proliferation of $(\Delta B,\Delta L)$ operators as a function of $d$, we show the number of non-derivative SMEFT ($\nu$SMEFT) operators in Fig.~\ref{fig:number_of_operators} (Fig.~\ref{fig:number_of_nuSMEFT_operators}). Here, operator really means field string, not counting flavor multiplicities or gauge and Lorentz contractions, so $LLHH$ counts as just 1 operator. Operators with $d$ that is divisible by 3 show the largest spread in $(\Delta B,\Delta L)$ since they can accommodate an additional fermion pair.
Overall, the number of operators that break baryon or lepton number grows exponentially, even when restricted to non-derivative one's.
Fig.~\ref{fig:dBdL} gives $d_\text{min}$  up to $d=25$ and \textit{vastly} extends previous maps~\cite{Weinberg:1980bf,Kobach:2016ami,Helset:2019eyc,Heeck:2019kgr} of the $\Delta B$/$\Delta L$ landscape.

Comparing to Fig.~\ref{fig:dBdL} shows that, surprisingly, Eq.~\eqref{eq:better_inequality} is an \textit{equality} for the $\nu$SMEFT up to  high $d$. As far as we can tell, it is only reduced to an inequality due to Pauli's exclusion principle~\cite{Pauli:1925nmn}, i.e.~the Grassmann nature of spinors, rarely taken into account in studies of $\Delta B$, $\Delta L$, and $d$. For example, each generation of $\nu$ can only arise \textit{twice} in a non-derivative operator, e.g.~written as $\bar{\nu}^C_{e}\nu_{e}$; any additional factor of $\nu_{e}$ would lead to a vanishing operator~\cite{Fonseca:2018aav}.
With three generations, the largest non-derivative $\nu$ product we can write down is $\bar{\nu}^C_{e}\nu_{e}\bar{\nu}^C_{\mu}\nu_{\mu}\bar{\nu}^C_{\tau}\nu_{\tau}$, which has $\Delta L=6$ and $d=9$. 
If we wanted to add \textit{additional} lepton number to this operator we could not simply multiply it by $\bar{\nu}^C_{\alpha}\nu_{\beta}$, since that would give zero following Pauli, nor by multiplying it with $D_\mu\bar{\nu}^C_{\alpha}\nu_{\beta}$, as that would violate $F$ parity. The only options are multiplication by $D_\mu^2\bar{\nu}^C_{\alpha}\nu_{\beta}$ or by the Weinberg operator $LLHH$; either way, the resulting $\Delta L = 8$ operator has $d=14$, two units more than the $d_\text{min}$ from Eq.~\eqref{eq:better_inequality}. Consequently, the equality in~\eqref{eq:better_inequality} only holds up to $d=9$ for for $\Delta L \neq 0 = \Delta B$ $\nu$SMEFT (much higher for $\Delta B\neq 0$ operators), but of course remains a correct inequality always. Concretely, Eq.~\eqref{eq:better_inequality} is an equality up to $d = 25$ except for $(|\Delta B|,|\Delta L|) = (0, > 6)$ and $(1,11)$ in the $\nu$SMEFT. 
While right-handed neutrino are not restricted by anomaly cancellation to come in three generations~\cite{Heeck:2012fw}, Pauli's exclusion principle eventually also becomes a problem for SM fields, albeit at much higher mass dimension. This means that the relationship between baryon \& lepton number and operator dimension is not just constrained by gauge and Lorentz invariance, but also by the finite number of fermion generations, an aspect that has not been appreciated so far.

While Eq.~\eqref{eq:better_inequality} is essentially an equality for all practical purposes in the $\nu$SMEFT, it is considerably weaker in the SMEFT. Indeed, $(\Delta B,\Delta L) = (0,2)$ gives $d_\text{min} \geq 3$, while the actual operator only arises at $d=5$~\cite{Weinberg:1979sa}, and a similar mismatch occurs for all $\Delta B=0 \neq \Delta L$ operators. The culprit at low $d$ is hypercharge invariance rather than Pauli, difficult to implement beyond $F$ parity~\cite{Kobach:2016ami}.  For $\Delta B \neq 0$ cases, Eq.~\eqref{eq:better_inequality} works well even in the SMEFT, satisfying the equality for $|\Delta L | <5$. In some cases, the actual minimal operator dimension is even 4 units above our lower bound, e.g.~$(\Delta B,\Delta L) = (1,7)$ arises at $d_\text{min}=19$ rather than $15$. Still, for all practical purposes, Eq.~\eqref{eq:better_inequality} is a SMEFT equality for $\Delta B\neq 0$.

In this letter, we revisited the connection between the mass dimension of effective operators and the baryon and lepton number they can carry. By combining Kobach's reasoning with Weinberg's $F$ parity, we find the improved inequality~\eqref{eq:better_inequality}. For the $\nu$SMEFT, this is a de-facto \textit{equality} for phenomenologically relevant operators, while the SMEFT shows a few outliers.
Our results shed additional light on the ($\nu$)SMEFT structure and highlight restrictions imposed by Pauli's exclusion principle.
A phenomenological study of high-$d$ operators carrying $\Delta B$ is currently under investigation~\cite{Heeck2025}.


\section*{Acknowledgements}
 We thank Ian Shoemaker for discussions. 
This work was supported in part by the U.S.~Department of Energy under Grant No.~DE-SC0007974  and a 4-VA at UVA Collaborative Research Grant.
We acknowledge Research Computing at The University of Virginia for providing computational resources that have contributed to the results reported within this publication.

\bibliographystyle{utcaps_mod}
\bibliography{biblio.bib}

\providecommand{\href}[2]{#2}\begingroup\raggedright\begin{thebibliography}{10}

\bibitem{Heeck:2019kgr}
J.~Heeck and V.~Takhistov, ``{Inclusive Nucleon Decay Searches as a Frontier of Baryon Number Violation},'' \href{http://dx.doi.org/10.1103/PhysRevD.101.015005}{{\em Phys. Rev. D} {\bfseries 101} (2020) 015005}, \href{http://arxiv.org/abs/1910.07647}{[{\ttfamily 1910.07647}]}.

\bibitem{Dev:2022jbf}
P.~S.~B. Dev {\em et~al.}, ``{Searches for baryon number violation in neutrino experiments: a white paper},'' \href{http://dx.doi.org/10.1088/1361-6471/ad1658}{{\em J. Phys. G} {\bfseries 51} (2024) 033001}, \href{http://arxiv.org/abs/2203.08771}{[{\ttfamily 2203.08771}]}.

\bibitem{FileviezPerez:2022ypk}
P.~F. Perez, A.~Pocar, K.~Babu, L.~J. Broussard, V.~Cirigliano, J.~Heeck, S.~Gardner, E.~Kearns, A.~J. Long, S.~Raby, R.~Ruiz, E.~Thomson, C.~E. Wagner, and M.~B. Wise, ``{On Baryon and Lepton Number Violation},'' \href{http://arxiv.org/abs/2208.00010}{[{\ttfamily 2208.00010}]}.

\bibitem{Broussard:2025opd}
L.~J. Broussard, A.~Crivellin, M.~Hoferichter, S.~Syritsyn, Y.~Aoki, J.~L. Barrow, A.~B. i~Beneito, Z.~Berezhiani, N.~F. Calabria, S.~Fajfer, S.~Gardner, J.~Heeck, C.~Jiang, L.~Naterop, A.~A. Petrov, R.~Shrock, A.~Thompson, U.~van Kolck, M.~L. Wagman, L.~Wan, J.~Womersley, and J.-S. Yoo, ``{Baryon Number Violation: From Nuclear Matrix Elements to BSM Physics},''
\newblock 2025.
\newblock \href{http://arxiv.org/abs/2504.16983}{[{\ttfamily 2504.16983}]}.

\bibitem{Weinberg:1979sa}
S.~Weinberg, ``{Baryon and Lepton Nonconserving Processes},'' \href{http://dx.doi.org/10.1103/PhysRevLett.43.1566}{{\em Phys. Rev. Lett.} {\bfseries 43} (1979) 1566--1570}.

\bibitem{Weinberg:1980bf}
S.~Weinberg, ``{Varieties of Baryon and Lepton Nonconservation},'' \href{http://dx.doi.org/10.1103/PhysRevD.22.1694}{{\em Phys. Rev. D} {\bfseries 22} (1980) 1694}.

\bibitem{Weldon:1980gi}
H.~A. Weldon and A.~Zee, ``{Operator Analysis of New Physics},'' \href{http://dx.doi.org/10.1016/0550-3213(80)90218-7}{{\em Nucl. Phys. B} {\bfseries 173} (1980) 269--290}.

\bibitem{Rao:1983sd}
S.~Rao and R.~E. Shrock, ``{Six Fermion ($B-L$) Violating Operators of Arbitrary Generational Structure},'' \href{http://dx.doi.org/10.1016/0550-3213(84)90365-1}{{\em Nucl. Phys. B} {\bfseries 232} (1984) 143--179}.

\bibitem{Degrande:2012wf}
C.~Degrande, N.~Greiner, W.~Kilian, O.~Mattelaer, H.~Mebane, T.~Stelzer, S.~Willenbrock, and C.~Zhang, ``{Effective Field Theory: A Modern Approach to Anomalous Couplings},'' \href{http://dx.doi.org/10.1016/j.aop.2013.04.016}{{\em Annals Phys.} {\bfseries 335} (2013) 21--32}, \href{http://arxiv.org/abs/1205.4231}{[{\ttfamily 1205.4231}]}.

\bibitem{deGouvea:2014lva}
A.~de~Gouvea, J.~Herrero-Garcia, and A.~Kobach, ``{Neutrino Masses, Grand Unification, and Baryon Number Violation},'' \href{http://dx.doi.org/10.1103/PhysRevD.90.016011}{{\em Phys. Rev. D} {\bfseries 90} (2014) 016011}, \href{http://arxiv.org/abs/1404.4057}{[{\ttfamily 1404.4057}]}.

\bibitem{Henning:2015alf}
B.~Henning, X.~Lu, T.~Melia, and H.~Murayama, ``{2, 84, 30, 993, 560, 15456, 11962, 261485, ...: Higher dimension operators in the SM EFT},'' \href{http://dx.doi.org/10.1007/JHEP08(2017)016}{{\em JHEP} {\bfseries 08} (2017) 016}, \href{http://arxiv.org/abs/1512.03433}{[{\ttfamily 1512.03433}]}. [Erratum: JHEP 09, 019 (2019)].

\bibitem{Kobach:2016ami}
A.~Kobach, ``{Baryon Number, Lepton Number, and Operator Dimension in the Standard Model},'' \href{http://dx.doi.org/10.1016/j.physletb.2016.05.050}{{\em Phys. Lett. B} {\bfseries 758} (2016) 455--457}, \href{http://arxiv.org/abs/1604.05726}{[{\ttfamily 1604.05726}]}.

\bibitem{Fonseca:2017lem}
R.~M. Fonseca, ``{The Sym2Int program: going from symmetries to interactions},'' \href{http://dx.doi.org/10.1088/1742-6596/873/1/012045}{{\em J. Phys. Conf. Ser.} {\bfseries 873} no.~1, (2017) 012045}, \href{http://arxiv.org/abs/1703.05221}{[{\ttfamily 1703.05221}]}.

\bibitem{Fonseca:2019yya}
R.~M. Fonseca, ``{Enumerating the operators of an effective field theory},'' \href{http://dx.doi.org/10.1103/PhysRevD.101.035040}{{\em Phys. Rev. D} {\bfseries 101} (2020) 035040}, \href{http://arxiv.org/abs/1907.12584}{[{\ttfamily 1907.12584}]}.

\bibitem{Helset:2019eyc}
A.~Helset and A.~Kobach, ``{Baryon Number, Lepton Number, and Operator Dimension in the SMEFT with Flavor Symmetries},'' \href{http://dx.doi.org/10.1016/j.physletb.2019.135132}{{\em Phys. Lett. B} {\bfseries 800} (2020) 135132}, \href{http://arxiv.org/abs/1909.05853}{[{\ttfamily 1909.05853}]}.

\bibitem{Pauli:1925nmn}
W.~Pauli, ``{\"Uber den Zusammenhang des Abschlusses der Elektronengruppen im Atom mit der Komplexstruktur der Spektren},'' \href{http://dx.doi.org/10.1007/BF02980631}{{\em Z. Phys.} {\bfseries 31} (1925) 765--783}.

\bibitem{Fonseca:2018aav}
R.~M. Fonseca and M.~Hirsch, ``{$\Delta L \ge 4$ lepton number violating processes},'' \href{http://dx.doi.org/10.1103/PhysRevD.98.015035}{{\em Phys. Rev. D} {\bfseries 98} (2018) 015035}, \href{http://arxiv.org/abs/1804.10545}{[{\ttfamily 1804.10545}]}.

\bibitem{Heeck:2012fw}
J.~Heeck, ``{Seesaw parametrization for $n$ right-handed neutrinos},'' \href{http://dx.doi.org/10.1103/PhysRevD.86.093023}{{\em Phys. Rev. D} {\bfseries 86} (2012) 093023}, \href{http://arxiv.org/abs/1207.5521}{[{\ttfamily 1207.5521}]}.

\bibitem{Heeck2025}
J.~Heeck, D.~Sokhashvili, and A.~Thapa, ``{Opening up baryon-number-violating operators}.'' To appear, 2025.

\end{thebibliography}\endgroup

\end{document}